\documentclass[aps,prb,twocolumn,showpacs,superscriptaddress,groupedaddress]{revtex4} 
\usepackage{amsmath,amssymb}
\usepackage{hyperref}
\usepackage{graphics,psfrag}
\usepackage{graphicx,psfrag}
\usepackage{epsfig}
\usepackage{longtable}
\usepackage{tabularx}
\usepackage{booktabs}
\usepackage[utf8x]{inputenc}
\usepackage[USenglish]{babel}
\usepackage{multirow}
\usepackage{rotating}
\usepackage{hhline}
\usepackage{xcolor}
\usepackage{ulem}

\begin{document}

\title{A study of Ti${}_n$O${}_{2n-1}$ Magnéli phases using Density Functional Theory}

\author{A. C. M. Padilha}
\email{antonio.padilha@ufabc.edu.br}
\affiliation{Centro de Ci\^encias Naturais e Humanas, Universidade Federal do ABC, Santo Andr\'e, SP, Brazil}

\author{J. M. Osorio-Guill\'en}
\affiliation{Instituto de F\'isica, Universidad de Antioquia UdeA, Calle 70 No. 52-51, Medell\'in, Colombia}
\affiliation{Centro de Ci\^encias Naturais e Humanas, Universidade Federal do ABC, Santo Andr\'e, SP, Brazil}

\author{A. R. Rocha}
\affiliation{Instituto de F\'isica Te\'orica, Universidade Estadual Paulista, S\~ao Paulo, SP, Brazil}

\author{G. M. Dalpian}
\email{gustavo.dalpian@ufabc.edu.br}
\affiliation{Centro de Ci\^encias Naturais e Humanas, Universidade Federal do ABC, Santo Andr\'e, SP, Brazil}

\date{\today}

% -----------------------------------------------------------------------------------------------------------

\begin{abstract}
  Defects in the rutile TiO${}_2$ structures have been extensively studied, but the intrinsic defects of the oxygen deficient Ti${}_n$O${}_{2n-1}$ phases have not been given the same amount of consideration. Those structures, known as Magn\'eli phases, are characterized by the presence of ordered planes of oxygen vacancies, also known as shear-planes, and it has been shown that they form conducting channels inside TiO-based memristor devices. Memristors are excellent candidates for a new generation of memory devices in the electronics industry. In this paper we present DFT-based electronic structure calculations for Ti${}_n$O${}_{2n-1}$ Magn\'eli structures using PBESol+U ($0 \leq \text{U} \leq 5$ eV) and HSE functionals, showing that intrinsic defects present in these structures are responsible for the appearence of states inside the bandgap, which can act as intrinsic dopants for the enhanced conductivity of TiO${}_2$ memristive devices.
\end{abstract}

\pacs{71.20.-b,71.20.Nr,71.55.-i}

\maketitle

% -----------------------------------------------------------------------------------------------------------

\section{Introduction}\label{sec:intro}

Resistance switching in oxides has been known for over half a century,\cite{Hickmott1962, Chopra1965, Argall1968, Johnson1968} but attention to this phenomena has received a boost since the realization of the memristor, an idea which was originally formulated by Chua in the 1970's\cite{chua_1971,chua_kang_1976}. Williams \textit{et al} showed, by manufacturing, measuring and switching the resistance state of such a device, basically a oxide thin film, that the memristance is a property that arises naturally at the nanoscale.\cite{S_Williams_nature_2008}

The working principle of the memristive-based memory device is the storage of information using the resistance state of a Metal-Insulator-Metal structure. This in turn can be changed (write operation) and measured (read) when subjected to an electric field. The insulator layer is a nanometer thick thin film which can be composed of a wide variety of materials, such as binary oxides,\cite{Szot2011} perovskites,\cite{Muenstermann2010,Szot2006,Beck2000,Rossel2001} as well as many other compounds which are known for their resistance switching properties.\cite{Waser2007,Hirose1976}

\begin{figure}[!ht]
 \centering
  \includegraphics[width=0.4\textwidth]{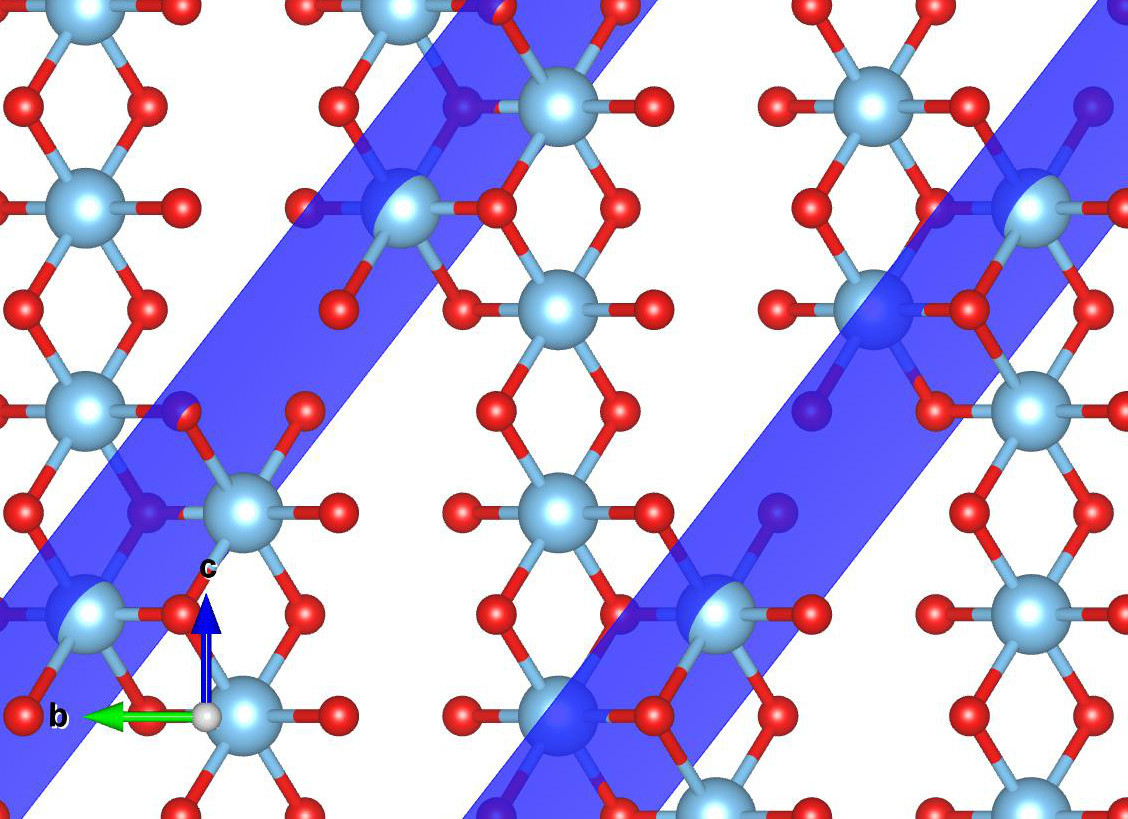}
  \caption{Cut of the Magn\'eli structure generated through the operation $(121)\frac{1}{2}[0\bar{1}1]$. The shear planes $(121)$ are featured in blue (online version).} 
  \label{fig:shear-plane}
\end{figure}

The atractiveness of these devices for memory storage resides in the fact that they would be faster, denser and less power consuming than those available today.\cite{Jeong2012} However, the mechanism for memristance is not, at present, well understood at the atomic level. Many authors point out to a phase transition taking place inside the oxide matrix for TiO${}_2$-based devices, leading to the formation of Magn\'eli-phases (Ti${}_n$O${}_{2n-1}$, $n =4,5$) conducting channels.\cite{Kwon2010,HwanKim2011} Those structures can be regarded as oxygen deficient TiO${}_2$, where the concentration of oxygen vacancies (V${}_{\mathrm{O}}$) is such that those defects become organized in a shear plane structure. The formation of these extended defects is exemplified by the operation $(121)\frac{1}{2}[0\bar{1}1]$ in the rutile structure, where the first three indices refer to a plane in the rutile structure and the last three to a displacement vector in the same structure.\cite{Andersson1960,Harada2010} An example of this structure is presented in Fig. \ref{fig:shear-plane}.

Despite the existence of many studies about defects in rutile TiO${}_2$, both experimental\cite{Henrich1976,Nolan2008,Gopel1984,Henderson2003,Yang2009,Wendt2008,Mitsuhara2012,Yim2010,Kruger2012} and theoretical,\cite{Nolan2008,Wendt2008} there are few works on the intrinsically defective Magn\'eli phases.\cite{Liborio2009a,Liborio2009b,Weissmann2011,Leonov2006,Guo2012,Bartholomew1969,Inglis1983,Clarke1988} Electronic structure calculations have been used to understand mainly the Ti${}_2$O${}_3$\cite{Guo2012} and Ti${}_4$O${}_7$.\cite{Liborio2009a,Liborio2009b,Weissmann2011,Leonov2006} Experimental reports\cite{Bartholomew1969,Inglis1983,Clarke1988} of some of these structures are not conclusive about the origin of the increase of electrical conductivity with respect to the rutile phase, neither about the switching mechanism. The question that arises is how the presence of extended-defect structures is related to the increase of electrical conductivity in this case.

Aiming to provide a general insight on the electronic structure of these oxygen deficient oxides - which are known to exist - we present a systematic study based on \textit{ab-initio} Density Functional Theory (DFT) electronic structure calculations for Ti${}_n$O${}_{2n-1}$, $2 \leq n \leq 5$. We used functionals based both on Generalized Gradient Approximation (GGA) as well as on a hybrid approach (including part of the Hartree-Fock exchange contribution). On-site Coulomb interaction (DFT+U) for electrons in Ti(d) orbitals was also introduced, since there are questions as to the validity of using hybrid functionals in Ti-based oxides\cite{Janotti2010}. We show that these oxygen defficient structures containing extended defects present localized states inside the gap, which in turn can act as intrinsic dopants to alter the transport properties of memristive devices. 

% -----------------------------------------------------------------------------------------------------------

\section{Simulation Details}

Simulations were performed using the Vienna Ab-Initio Software Package - VASP\cite{Kresse1993,Kresse1994,Kresse1996,KresseFurthmuller1996} - within the Projector Augmented Waves scheme.\cite{Blochl1994,Kresse1999} The GGA functional PBESol\cite{PBESol} - with and without on-site Coulomb interaction (within Dudarev's Approach\cite{Dudarev}) for Ti(d) orbitals - and hybrid HSE\cite{HSE} (20\% HF exchange) were used for ionic relaxations (forces $< 2.5 \times 10^{-3}$ eV / \AA) and orbital projected density of states (PDOS) calculations. The values of U (in fact, it only matters U-J when one uses this implementation, hence, for all calculations, we used J = 0) used for the on-site interaction ranged from 0 to 5 eV. For all structures spin polarization was taken into account. Monkhorst-Pack sampling was used for ionic relaxation with the GGA functional and $\Gamma$-centered meshes for PDOS calculations after that, using an energy cutoff of 520 eV. For the hybrid functional calculations, only $\Gamma$-centered meshes were used for both relaxation and PDOS, with a smaller cutoff (400 eV), in such a manner that the k-points would be more equally spaced in all directions. Finally the 3p3d4s and 2s2p configurations were considered as valence electrons for Ti and O atoms respectively.

The use of the Hubbard U parameter resulted in a change of the electronic structure when the system was driven from the delocalized regime, (GGA functional calculation, U = 0) that neglects the orbital dependence of the Coulomb interaction, to the regime where orbital dependence is included, (U $>$ 0) leading to a better description of localized states (Ti d orbitals). We used values for U that ranged from 0 to 5 eV, but we report only the meaningful results for U = 0 and 5 eV.
 
When possible, unit cell symmetry was used to obtain the primitive unit cell from crystallographic data using the Phonopy software.\cite{Phonopy} All crystal structure images were generated using the Vesta software.\cite{Vesta}

% -----------------------------------------------------------------------------------------------------------

\section{Results}

\subsection{Structural Properties}

Table \ref{tab:symm} lists the space groups for all structures studied in this work. From X-ray diffraction experiments,\cite{Robinson1974,Abrahams1963} it is known that Ti${}_2$O${}_3$ presents a rhombohedral corundrum-like structure with space group $R\bar{3}c$ where the Ti atoms are enclosed by oxygen in octahedral format. These polyhedra are the building blocks of this structure, as well as all the other structures presented here, and hereafter will be refered to as TiO${}_6$ octahedra. For Ti${}_2$O${}_3$, these building blocks are displaced in face-sharing pairs. 
\begin{center}
 \begin{table}[h!]
  \centering
   \caption{\label{tab:symm} Space group in the International and Schoenfiles notations for all Ti${}_n$O${}_{2n-1}$, ($2 \leq n \leq 5$ for the structures used in this work) after ionic relaxation.}
   \begin{tabular}{*{3}{p{0.3\columnwidth}}} 
   \hhline{===}
                              & \multicolumn{2}{c}{Space Group}    \\
                              & International (\#) &  Schoenflies  \\
    \hline
    Ti${}_2$O${}_3$           & $R\bar{3}c$ (167)  & $D_{3d}$      \\
    $\alpha$-Ti${}_3$O${}_5$  & $Cmcm$ (63)        & $D_{2h}$      \\
    $\beta$-Ti${}_3$O${}_5$   & $C2/m$ (12)        & $C_{2h}$      \\
    $\gamma$-Ti${}_3$O${}_5$  & $C2/c$ (15)        & $C_{2h}$      \\
    Ti${}_4$O${}_7$           & $P\bar{1}$ (2)     & $C_i$         \\
    Ti${}_5$O${}_9$           & $P1$ (1)           & $C_1$  \\ 
    \hhline{===}
   \end{tabular}
 \end{table}
\end{center}

The Ti${}_3$O${}_5$ structure obtained from crystallographic databases was not unique. The three phases, $\alpha$ (orthorhombic, anosovite-like, group $Cmcm$),\cite{Rusakov2002} $\beta$ (monoclinic, group $C2/m$)\cite{Asbrink:a02589,Grey1994} and $\gamma$-Ti${}_3$O${}_5$ (monoclinic, $I2/c$)\cite{Hong:a21519} were used for the calculations. Onoda \textit{et al.} pointed out a first-order phase transition at approximately 440 K - 460 K \cite{Onoda1998} between $\alpha$ and $\beta$-Ti${}_3$O${}_5$. The transition from $\beta$ to the room-temperature $\gamma$ phase at $\approx$ 250 K was reported by Hong and \AA sbrink.\cite{Hong:a21519}

Ti${}_4$O${}_7$ presents a Magn\'eli structure composed, as described in the other structures, by TiO$_6$ octahedra for all the three phases characterized by Marezio \textit{et al.}.\cite{LePage1984,Marezio1973,Marezio1971} This structure can be viewed as infinite planes of rutile, $n$-TiO${}_6$ octahedra thick along the (121) direction of the rutile crystal, limited by a plane of defects (oxygen vacancies), which characterizes a crystallographic shear structure.\cite{Harada2010,Liborio2009a} The unit cells obtained from crystallographic databases was triclinic and presented space group $P\bar{1}$ for the three known phases: High (HT), Intermediate (IT), and Low-temperature (LT). The only difference between those structures was a slight displacement of the atoms, whose positions became essencially equal after ionic relaxation. This also has lead to identical PDOS in all cases. Because of that, we present only the results for the relaxation using the LT phase as a starting configuration in Table \ref{tab:struct}.

Finally, Ti${}_5$O${}_9$ belongs to the same Magn\'eli series (Ti${}_n$O${}_{2n-1}$), presenting a triclinic unit cell. The difference is that the rutile-like planes are one extra unit of TiO${}_6$ thicker. The initial structure used for our calculations was obtained by Andersson.\cite{Andersson1960}

The structural parameters, after relaxation, for all oxides studied in this work are listed on Table \ref{tab:struct}. The difference beteween experimental and calculated lattice paremeters was lower than 6.0\% for all structures and both functionals, and it was not possible to notice any systematic underestimation or overestimation of the values regarding the functional chosen for the simulations. 
 \begin{table*}[Ht!]
  \centering
  \caption{\label{tab:struct} Experimental and theoretical values of lattice parameters for the Ti${}_2$O${}_3$, $\alpha$, $\beta$ and $\gamma$-Ti${}_3$O${}_5$, Ti${}_4$O${}_7$, and Ti${}_5$O${}_9$ structures. Mean Absolute Relative Error for unit cell volume is also presented in parenthesis. Values presented are those of the primitive cells.}
   \begin{tabular}{*{2}{p{0.1\columnwidth}} *{7}{p{0.19\columnwidth}} *{1}{p{0.25\columnwidth}}} 
	\hhline{==========}
       &               &                          & $a$(\AA) & $b$(\AA) & $c$(\AA) & $\alpha$(${}^{\mathrm{0}}$) & $\beta$(${}^{\mathrm{o}}$) & $\gamma$(${}^{\mathrm{o}}$) & $\Omega$ (\AA${}^3$) \\
	\hhline{==========}
  &  &    Exp\cite{Robinson1974,Abrahams1963} & 5.433	& -      & -        & 56.57  & -      & -                         & 160.37               \\
  &  &    PBESol                              & 5.471   & -      & -        & 54.89  & -      & -                         & 163.80 (2.14\%)      \\
  \multicolumn{2}{c}{\multirow{-3}{*}{\begin{sideways}Ti${}_2$O${}_3$\end{sideways}}} &  HSE           & 5.370  & -      & -        & 57.62   & -      & -               & 154.87 (3.43\%)       \\
             \hhline{==========}
       &               & Exp\cite{Rusakov2002}    & 3.747    & 5.090    & 9.715    & 90.00                       & 90.00                      & 68.40                       & 172.27 \\
       &      $\alpha$ & PBESol                   & 3.760    & 5.237    & 9.937    & 90.00                       & 90.00                      & 68.96                       & 182.60 (6.00\%) \\
       &               & HSE                      & 3.682    & 5.258    & 9.978    & 90.00                       & 90.00                      & 69.50                       & 180.94 (5.03\%) \\
             \cline{2-10}
       &               & Exp\cite{Asbrink:a02589} & 3.802    & 5.233    & 9.442    & 91.79                       & 90.00                      & 111.30                      & 174.94 \\
       &      $\beta$  & PBESol                   & 3.834    & 5.195    & 9.215    & 90.87                       & 90.00                      & 111.65                      & 170.60 (2.48\%) \\
       &               & HSE                      & 3.791    & 5.196    & 9.173    & 90.76                       & 90.00                      & 111.40                      & 168.23 (3.61\%) \\
             \cline{2-10}
      &                & Exp\cite{Hong:a21519}    & 5.075    & 5.658    & 7.181    & 109.58                      & 90.00                      & 116.64                      & 170.85 \\
     &        $\gamma$ & PBESol                   & 4.997    & 5.627    & 7.180    & 109.81                      & 90.00                      & 116.36                      & 167.45 (1.99\%) \\
   \multirow{-9}{*}{\begin{sideways}Ti${}_3$O${}_5$\end{sideways}} &                  & HSE                      & 5.076    & 5.664    & 7.069    & 109.36                      & 90.00                      & 116.62                      & 168.76 (1.22\%) \\
	\hhline{==========}
 &  &   Exp\cite{LePage1984} & 5.626    & 6.892    & 7.202    & 63.71                       & 109.68                     & 105.24                      & 233.60              \\
 &  &  PBESol               & 5.569    & 6.868    & 7.092    & 64.22                       & 109.72                     & 104.91                      & 229.12 (1.92\%)            \\
 \multicolumn{2}{c}{\multirow{-3}{*}{\begin{sideways}Ti${}_4$O${}_7$\end{sideways}}}  &  HSE                  & 5.618    & 6.898    & 7.076    & 63.77                       & 108.77                     & 104.23                      & 231.09 (1.07\%) \\
	\hhline{==========}
 &  &    Exp\cite{Andersson1960} & 5.569    & 7.120    & 8.865    & 97.55                       & 112.34                     & 108.50                      & 295.33              \\
 &  &   PBESol                  & 5.558    & 7.110    & 8.846    & 97.75                       & 112.51                     & 108.61                      & 292.54 (0.94\%)             \\
 \multicolumn{2}{c}{\multirow{-3}{*}{\begin{sideways}Ti${}_5$O${}_9$\end{sideways}}} &   HSE                     & 5.550    & 7.040    & 8.763    & 96.96                       & 112.35                     & 108.09                      & 289.68 (1.91\%) \\
	\hhline{==========}
   \end{tabular}
 \end{table*}

% -----------------------------------------------------------------------------------------------------------

\subsection{Electronic and Magnetic Properties}

Orbital-projected Density of States (PDOS) and magnetization profiles ($\mu(\vec{r})= \rho_{\uparrow}(\vec{r})-\rho_{\downarrow}(\vec{r})$) were obtained for all systems using both PBESol+U ($0 \leq \text{U} \leq 5$ eV) and HSE functionals. It is a characteristic of GGA functionals to underestimate the band gap, while calculations using hybrid functionals frequently result in better agreement with experimental data for some oxides.\cite{DiValentin2006} The use of the Hubbard U parameter presents a better description of the localization of d orbital electrons in transition metals oxides which is exactly the case of the structures studied in this work. 

The positioning of Ti${}_n$O${}_{2n-1}$ defect levels is a key ingredient to determine the electrical conductivity - specially its enhancement in memristive devices - of these structures. Thus we used both methodologies to study the electronic structure of these defect levels.

\begin{figure}[!ht]
 \centering
  \includegraphics[width=0.5\textwidth]{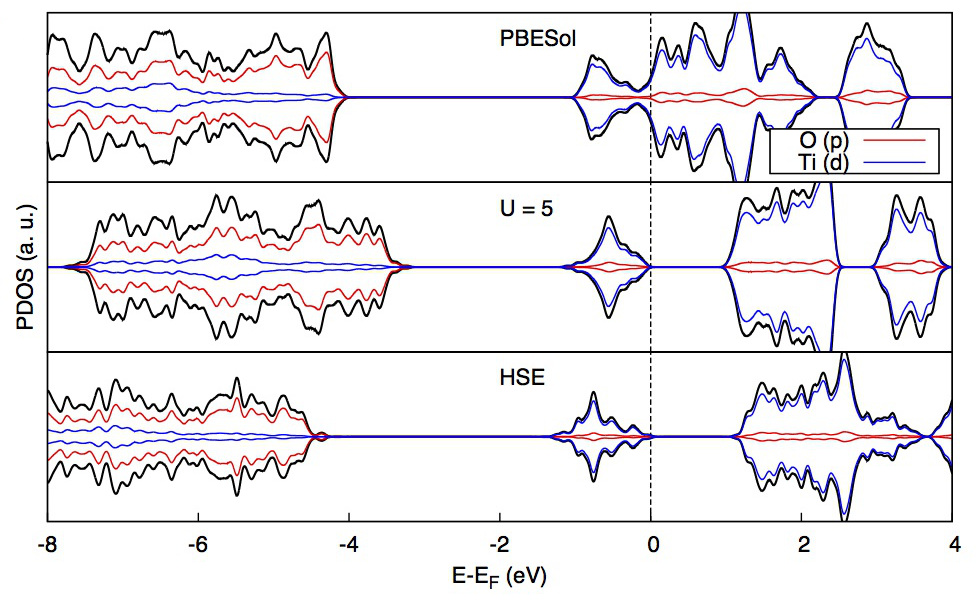}
  \caption{Ti${}_2$O${}_3$ PDOS obtained with PBESol functional (upper panel), PBESol+U (middle panel), and HSE (lower panel).} 
  \label{fig:dos-ti2o3}
\end{figure}

The PBESol calculations resulted in a metallic behavior. It is evident from the Ti${}_2$O${}_3$ PDOS in Fig. \ref{fig:dos-ti2o3} that some states, mainly composed of Ti(d) orbitals, move away from the unoccupied levels as the parameter U is increased, reaching the same qualitative results as in the HSE case for U = 5. This effect was expected owing to the better description, compared to GGA, of d orbitals by hybrid functionals\cite{Tran2006} as well as the more localized character of these orbitals with respect to increasing U. These levels can be interpreted as a defect-like level inherent to the crystalline structure, which can be split from the other Ti(d) states that are unoccupied with increasing U.

Experimental $E_g$ for this structure is 1 eV,\cite{Uchida2008} which is in excellent agrement with the results using PBESol+U, U = 5 eV, and HSE, given that the bandgap is interpreted as the energy difference between the defect levels and the CBM in Fig. \ref{fig:dos-ti2o3}. 

\begin{figure}[!ht]
 \centering
  \includegraphics[width=0.5\textwidth]{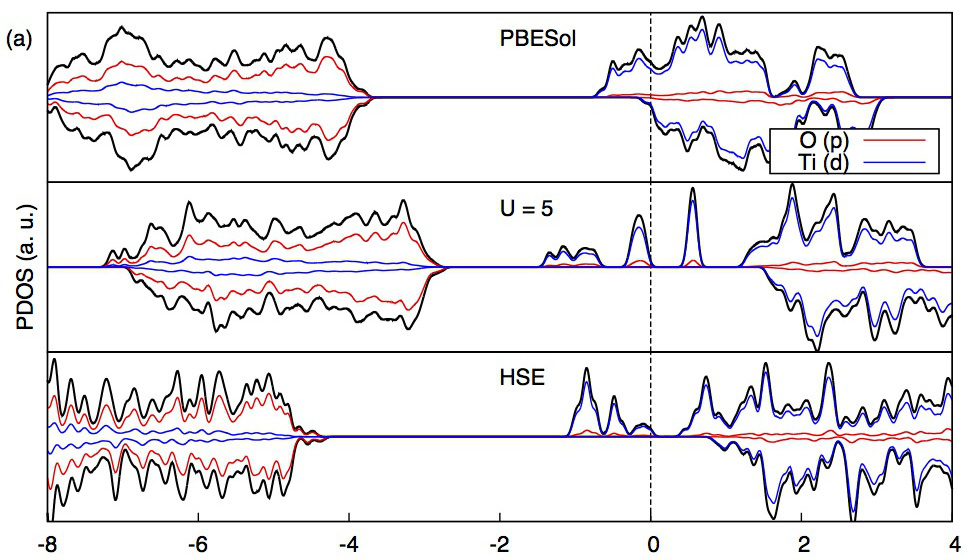} \\
  \includegraphics[width=0.5\textwidth]{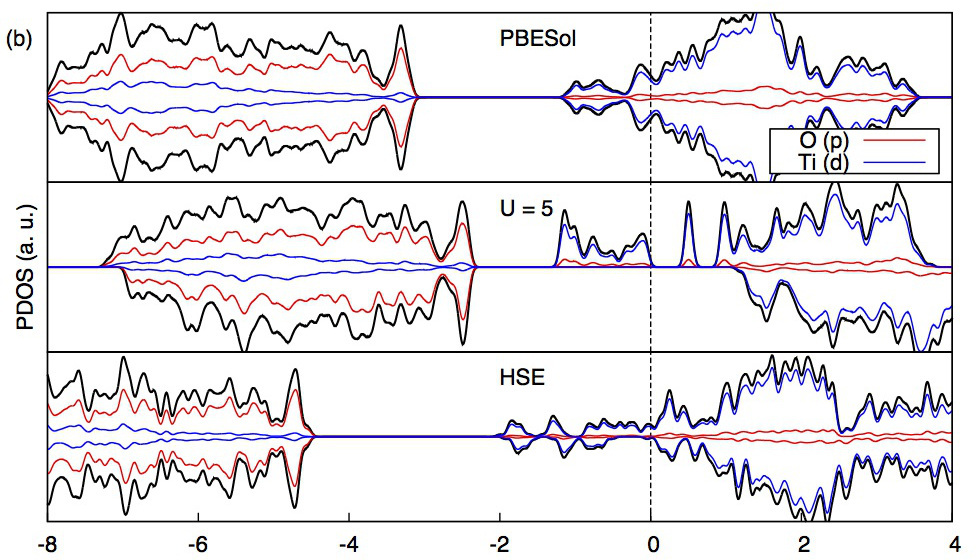} \\
  \includegraphics[width=0.5\textwidth]{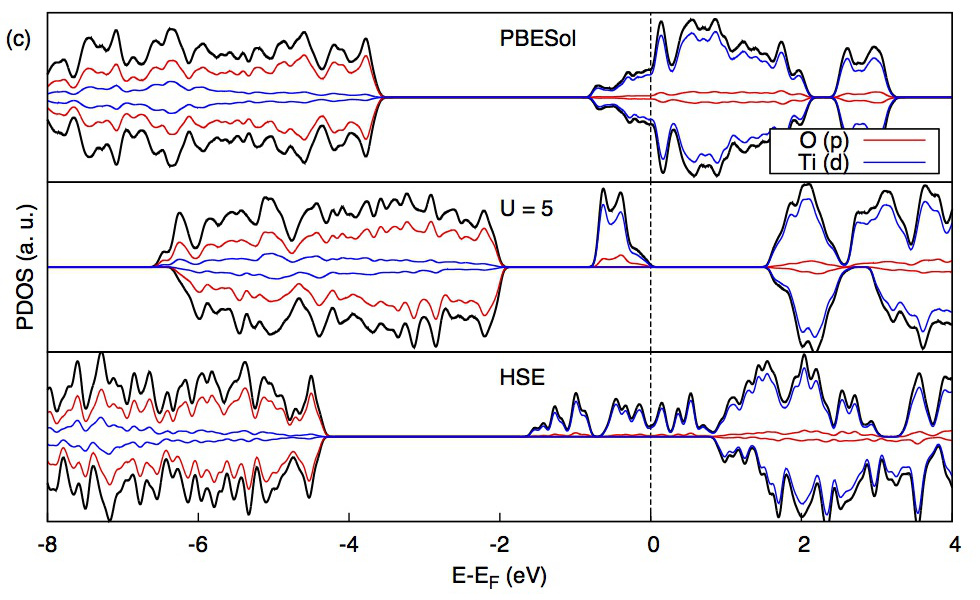} 
  \caption{$\alpha$- (a), $\beta$- (b) and $\gamma$-Ti${}_3$O${}_5$ (c) PDOS obtained with PBESol functional (upper panel), PBESol+U (middle panel), and HSE (lower panel).} 
  \label{fig:dos-ti3o5}
\end{figure}

In the case of Ti${}_3$O${}_5$, some states - mainly of Ti(d) character - were split from the unoccupied levels, leading to the formation of defect levels inside the bandgap when used U = 5, while for HSE calculations, some states remained very close to the Fermi energy, leading to a metallic like character for $\beta$- and $\gamma$-Ti${}_3$O${}_5$ (Fig. \ref{fig:dos-ti3o5}). According to Rao \textit{et al}\cite{Rao197183} and Bartholomew and Frankl,\cite{Bartholomew1969} this system is a semiconductor for temperatures below $\approx$ 400 K, therefore the description of the room-temperature phase $\gamma$-Ti${}_3$O${}_5$ as a metal by HSE is by no means correct.

\begin{figure}[!ht]
 \centering
  \includegraphics[width=0.5\textwidth]{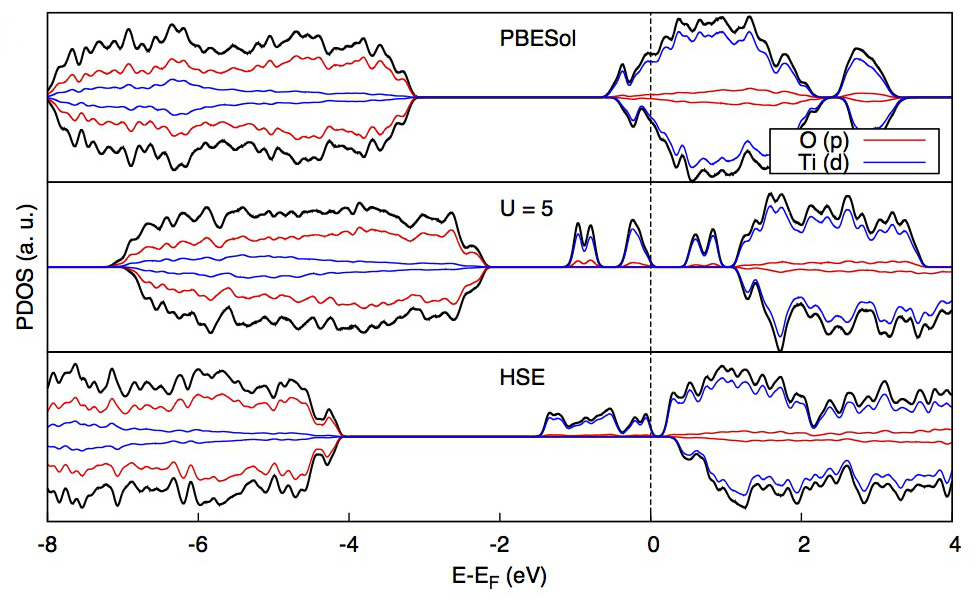}
  \caption{Ti${}_4$O${}_7$ PDOS obtained with PBESol functional (upper panel), PBESol+U (middle panel), and HSE (lower panel).} 
  \label{fig:dos-ti4o7}
\end{figure}

As previously discussed, total energy, PDOS (Fig. \ref{fig:dos-ti4o7}) and magnetization  were essentially equal for LT, IT and HT-Ti${}_4$O${}_7$ calculations using PBESol functional. This was expected since DFT calculations always take place at $T = 0$K leading the system to relax to the LT-structure in all three cases. Calculations with HSE showed that the magnetic ordering is dependent on initial conditions, although it does not influence the final atomic arrangement. The presence of magnetization in these structures leads to a lower energy configuration, a fact that was confirmed with calculations without spin polarization, which resulted in higher total energies.

\begin{figure}[!ht]
 \centering
  \includegraphics[width=0.5\textwidth]{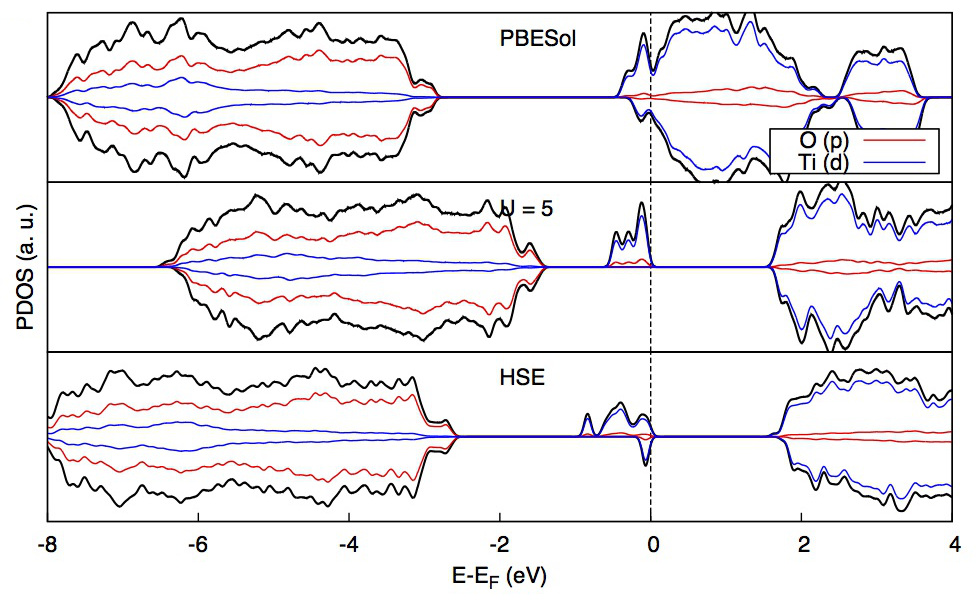}
  \caption{Ti${}_5$O${}_9$ PDOS obtained with PBESol functional (upper panel), PBESol+U (middle panel), and HSE (lower panel).} 
  \label{fig:dos-ti5o9}
\end{figure}

Similar results were obtained for Ti${}_5$O${}_9$ (Fig. \ref{fig:dos-ti5o9}), but one striking difference was the depth of the defect-like levels in this case: about 1.8 eV lower than the CBM with the HSE functional. It is also worth pointing out that for this system, HSE gives different descriptions for Ti${}_4$O${}_7$ and Ti${}_5$O${}_9$. The first seems to be a half metal whereas the former does not.

The main result used to decide which methodology between hybrid approach and the use of the Hubbard U parameter is better suited to study these systems is the agreement of the bandgap obtained through the calculations with the experimental one. This task, which in principle seems quite straightforward, is not simple for the two structures of interest for memristor applications, the Ti${}_4$O${}_7$ or the Ti${}_5$O${}_9$. For the former, there is no conclusive experimental data on the bandgap energy and for the latter, as to our knowledge, no reports at all. The bandgap of Ti${}_4$O${}_7$ is reported to be 0.041 eV from conductivity measurements\cite{Mulay1970}, 0.6 eV from spectroscopy data\cite{Abbate1995}, and 0.25 eV from optical transmission data\cite{Kaplan1977}. In principle, both HSE and U = 5 result in compatible bandgaps. Although the DFT+U is not designed to increase the bandgap, it is a consequence of the localization of the d orbitals when this methodology is used. 

%The fact that both methodologies give compatible results is not a proof that those results are correct. An example where both DFT+U and hybrid approaches fail is the case of the energy ordering of MnO polymorphs, where it was necessary to introduce an external potential acting only on Mn(d) orbitals to correct the underestimated exchange splitting of those levels\cite{Peng2013}. In that work, GW calculations were used as a benchmark, while in this present work, due to computational cost, we did not follow that path, but it could be an interesting study for future works.

 \begin{table*}[Ht!]
  \centering
  \caption{\label{tab:mags} Total magnetization per unit cell (Ti atom) in units of Bohr magnetons ($\mu_B$) and respective magnetization density ($\mu(\vec{r}) = \rho_{\uparrow}(\vec{r})-\rho_{\downarrow}(\vec{r})$) plots over the unit cell for all Ti${}_n$O${}_{2n-1}$ structures presented in this work obtained with PBESol, U = 5, and HSE functionals.}
   \begin{tabular}{p{0.1\columnwidth} *{6}{c}}
	\hhline{=======} 
                   & \multirow{2}{*}{Ti${}_2$O${}_3$} & \multicolumn{3}{c}{Ti${}_3$O${}_5$}   & \multirow{2}{*}{Ti${}_4$O${}_7$} & \multirow{2}{*}{Ti${}_5$O${}_9$} \\
	 &        & \hhline{---} & & \\
		   &                                  & $\alpha$ & $\beta$         & $\gamma$ &                                  &                                  \\
    \hline
	 & \includegraphics[height=1.8cm,trim=0 0 0 -5]{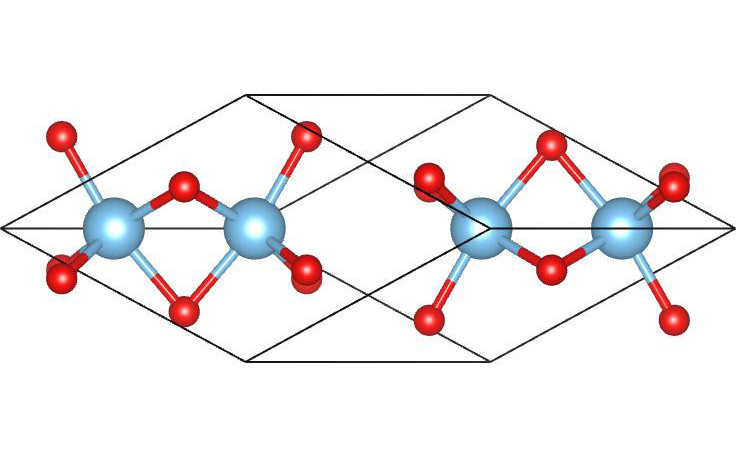} & \includegraphics[height=1.8cm,trim=0 0 0 -5]{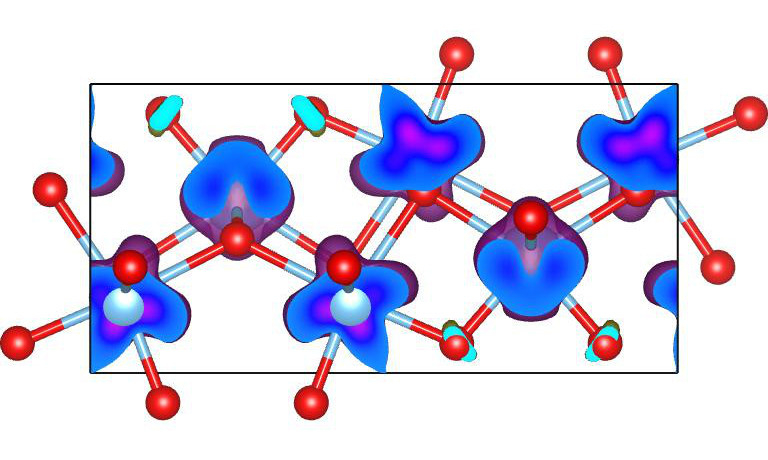} & \includegraphics[height=1.8cm,trim=0 0 0 -5]{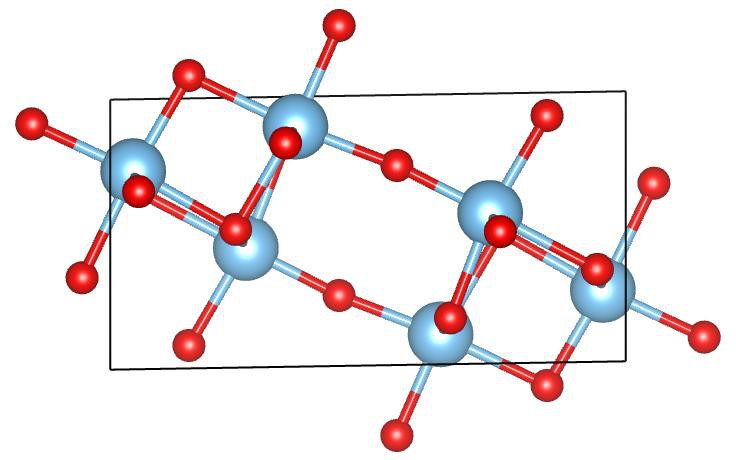} & \includegraphics[height=1.8cm,trim=0 0 0 -5]{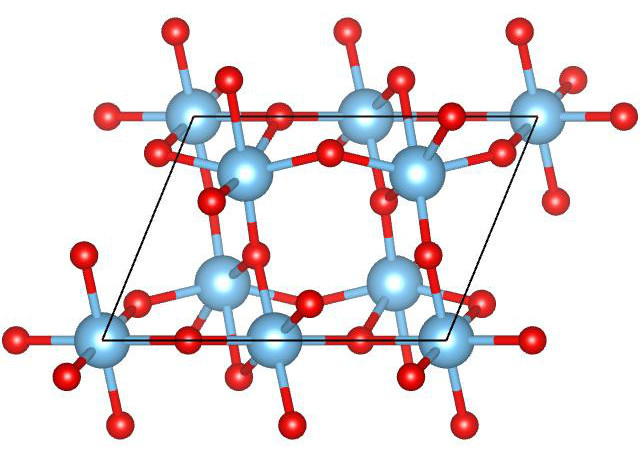} & \includegraphics[height=1.8cm,trim=0 0 0 -5]{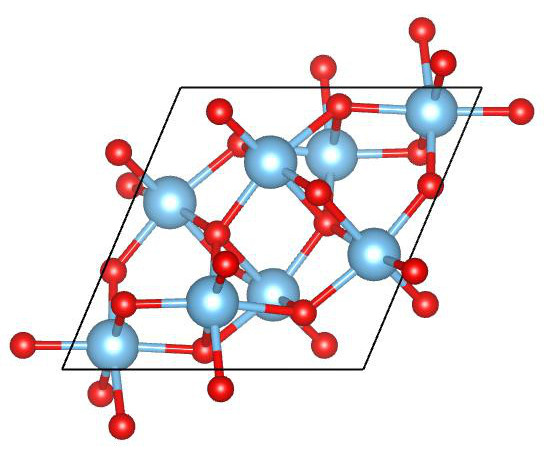} & \includegraphics[height=1.8cm,trim=0 0 0 -5]{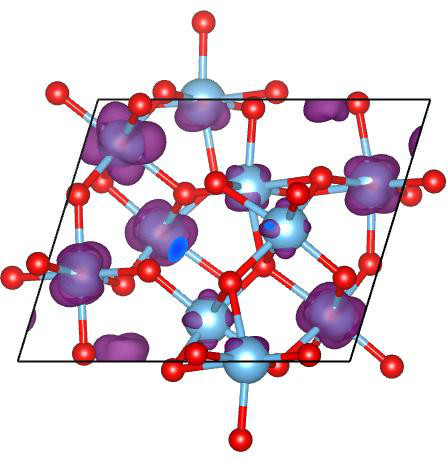} \\
     \multirow{-2}{*}[10mm]{\begin{sideways}PBESol\end{sideways}} & -0.06 (-0.01)                      & 3.75 (0.62)    & 0.00 (0.00)     & 0.15 (0.02)     & 1.12 (0.14)                            & 2.19 (0.22)                            \\
    \hline
	& \includegraphics[height=1.8cm,trim=0 0 0 -5]{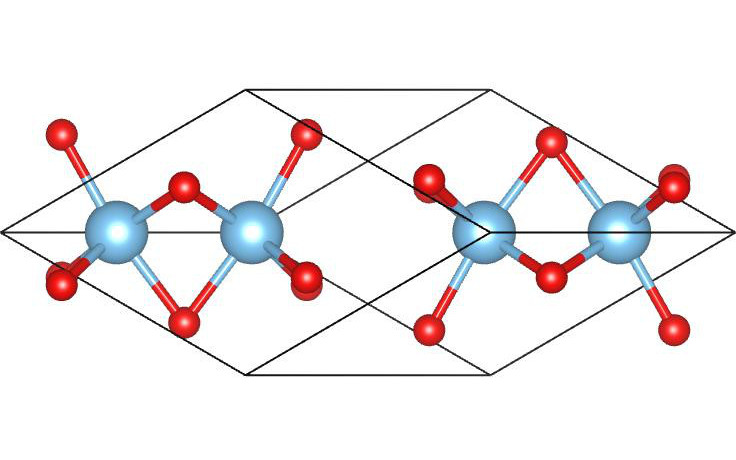} & \includegraphics[height=1.8cm,trim=0 0 0 -5]{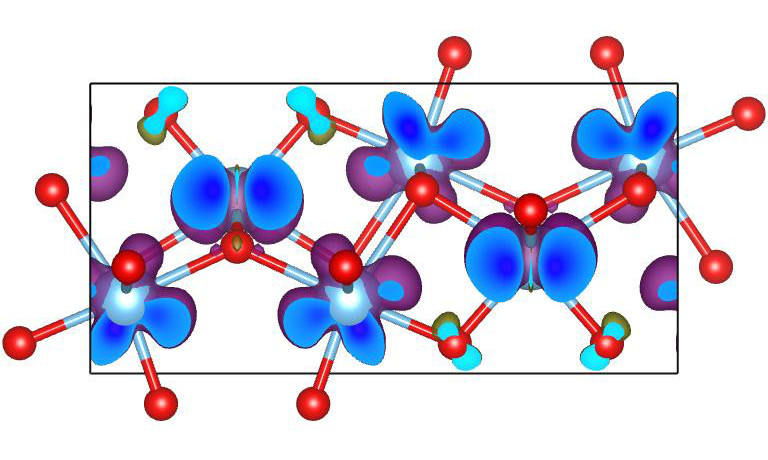} & \includegraphics[height=1.8cm,trim=0 0 0 -5]{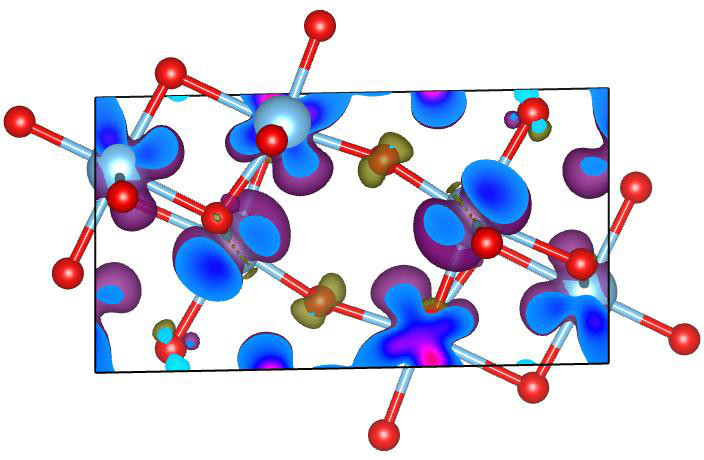} & \includegraphics[height=1.8cm,trim=0 0 0 -5]{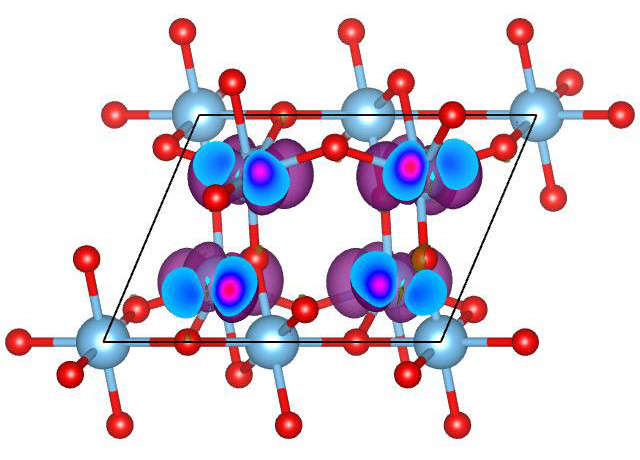} & \includegraphics[height=1.8cm,trim=0 0 0 -5]{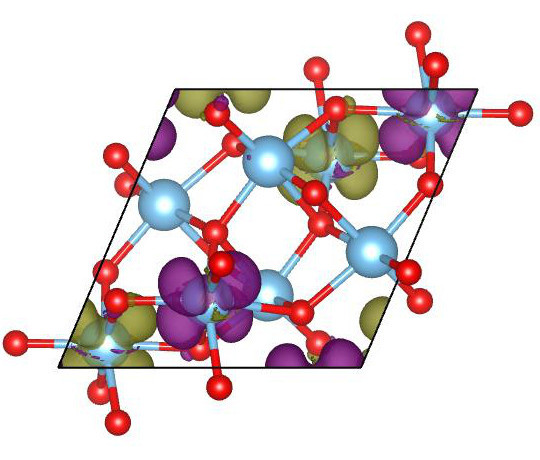} & \includegraphics[height=1.8cm,trim=0 0 0 -5]{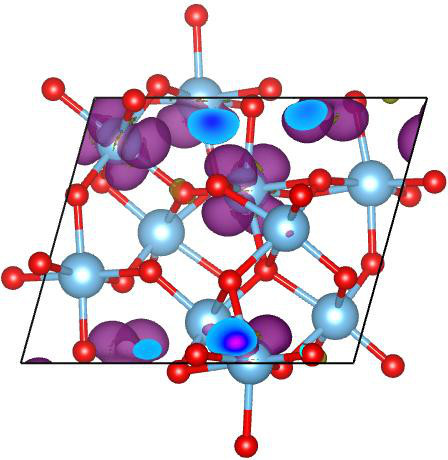} \\
     \multirow{-2}{*}[10mm]{\begin{sideways}U = 5\end{sideways}} &  0.00 (0.00)                           & 4.00 (0.67)    & 3.99 (0.66)           & 4.00 (0.67)    & 4.00 (0.50)                            & 4.00 (0.40)                            \\
    \hline
	& \includegraphics[height=1.8cm,trim=0 0 0 -5]{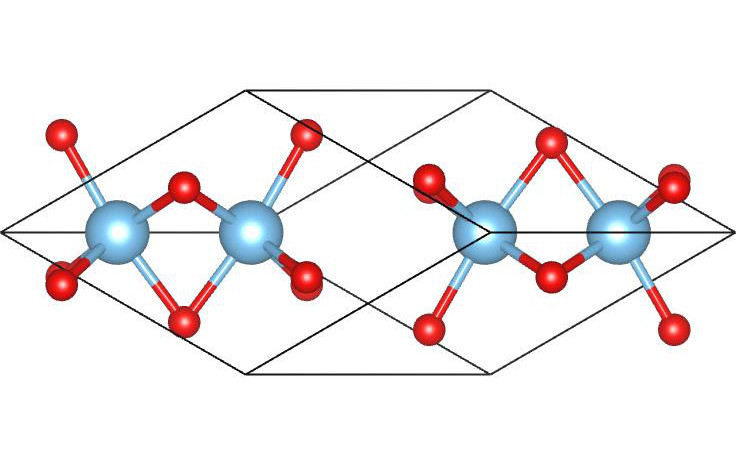} & \includegraphics[height=1.8cm,trim=0 0 0 -5]{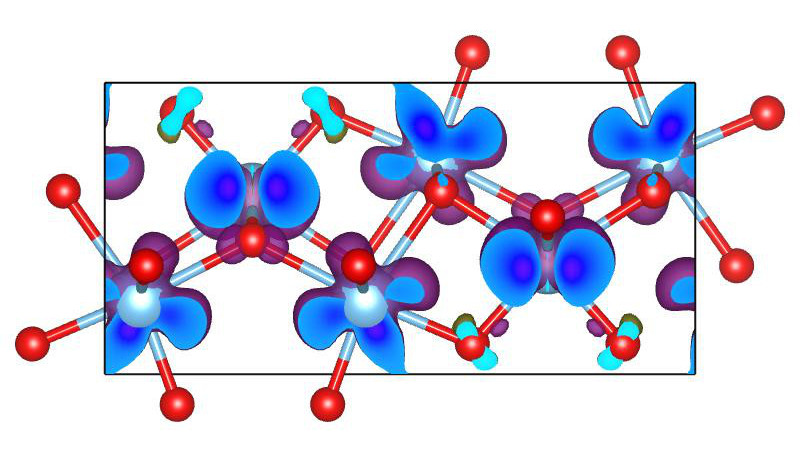} & \includegraphics[height=1.8cm,trim=0 0 0 -5]{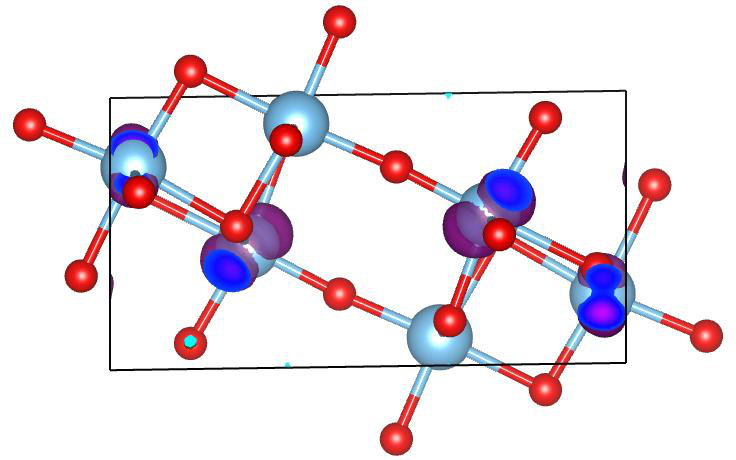} & \includegraphics[height=1.8cm,trim=0 0 0 -5]{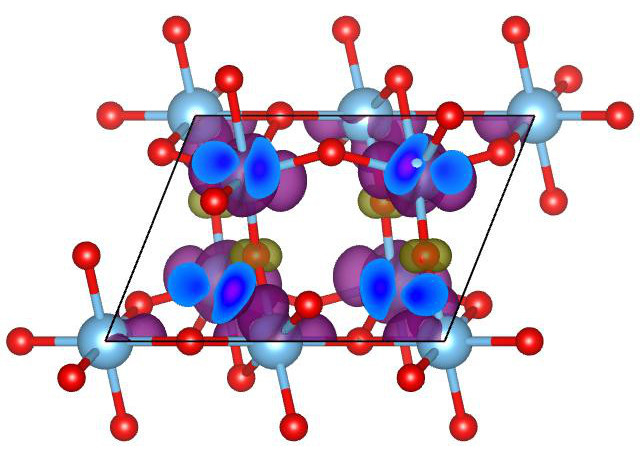} & \includegraphics[height=1.8cm,trim=0 0 0 -5]{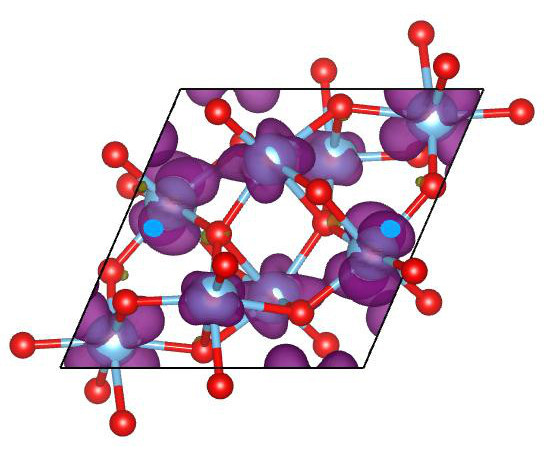} & \includegraphics[height=1.8cm,trim=0 0 0 -5]{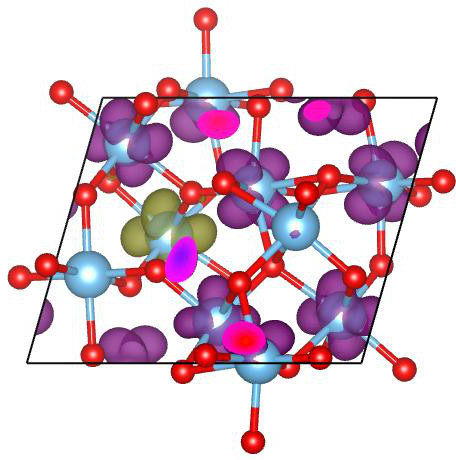} \\
     \multirow{-2}{*}[10mm]{\begin{sideways}HSE\end{sideways}} &  0.00 (0.00)                           & 4.00 (0.67)    & 0.94 (0.16)  & 4.00 (0.67)    & 4.00 (0.50)                            & 2.75 (0.27)                         \\ 
	\hhline{=======} 
   \end{tabular}
 \end{table*}

 \begin{table*}[Ht!]
  \centering
  \caption{\label{tab:parchg} Band decomposed chargedensity plots over the unit cell over the defect levels for Ti${}_2$O${}_3$ and Ti${}_4$O${}_7$ using PBESol, PBESol+U (U = 5 eV) and HSE functionals.}
   \begin{tabular}{p{0.1\columnwidth} *{3}{c}}
	\hhline{====}
	   & PBESol & U = 5 & HSE \\
      \multirow{-1}{*}[12mm]{\begin{sideways}Ti${}_2$O${}_3$\end{sideways}} & \includegraphics[width=0.6\columnwidth]{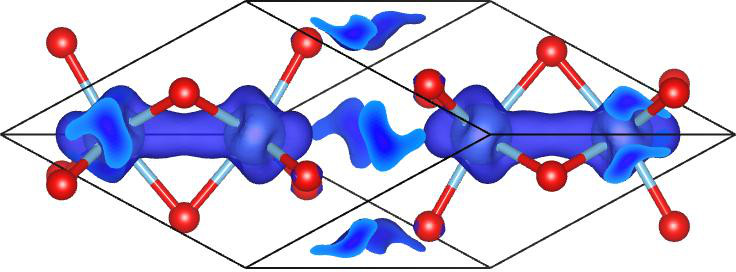} & \includegraphics[width=0.6\columnwidth,trim = 0 0 0 -5]{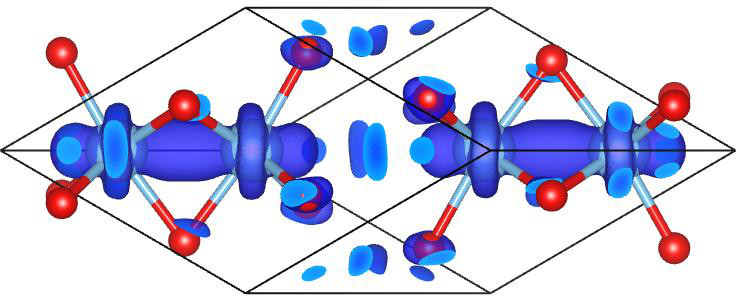}  & \includegraphics[width=0.6\columnwidth]{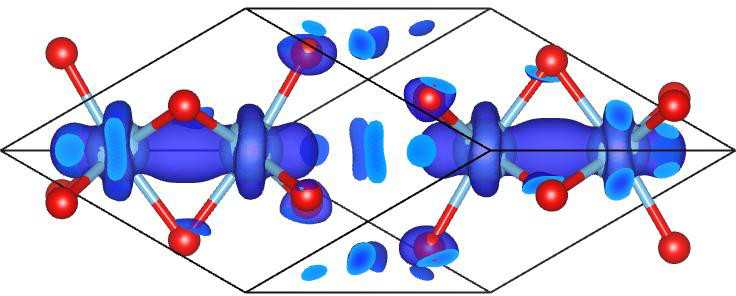} \\
	\hhline{----} 
      \multirow{-1}{*}[22mm]{\begin{sideways}Ti${}_4$O${}_7$\end{sideways}} & \includegraphics[width=0.6\columnwidth]{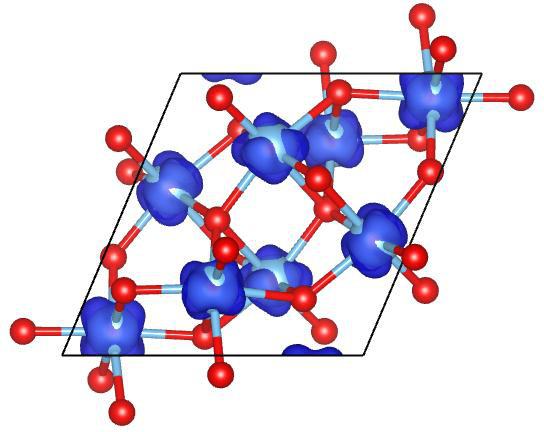} & \includegraphics[width=0.6\columnwidth,trim = 0 0 0 -5]{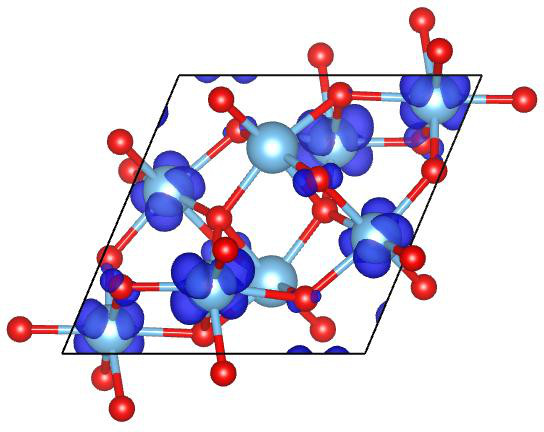}  & \includegraphics[width=0.6\columnwidth]{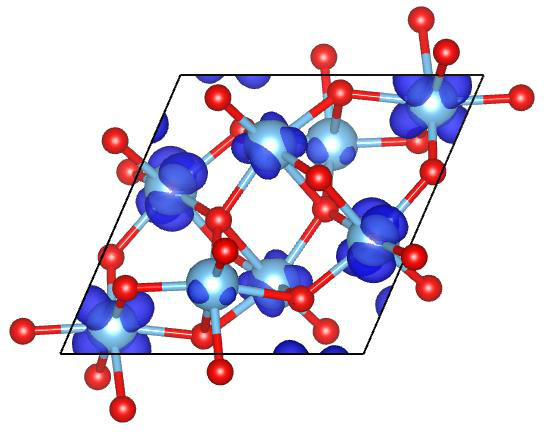} \\
	\hhline{====} 
   \end{tabular}
 \end{table*}

 The magnetization for all structures was higher for U = 5 eV as well as HSE functional calculations compared to GGA alone, with the exception of Ti${}_2$O${}_3$. These results are shown in Table \ref{tab:mags}. As is known for the same type of defects (V${}_{\mathrm{O}}$'s) in the TiO${}_2$ rutile, the removal of oxygen atoms from the oxide structure leaves a pair of unbound electrons.\cite{Lee2012,Janotti2010} Each of the Ti${}_n$O${}_{2n-1}$ phases presents two V${}_{\mathrm{O}}$'s per unit cell, therefore there are four electrons left behind. Magnetization density plots for Ti${}_n$O${}_{2n-1}$, $2 \leq n \leq 5$ are also shown in Table \ref{tab:mags}, where it is possible to identify these electrons laying on the Ti(d) orbitals through the unit cell. They are also responsible for the defect levels inside the bandgap seen in all structures. 
 
 The band decomposed charge density for both Ti${}_2$O${}_3$ and Ti${}_4$O${}_7$ defect levels are presented in Table \ref{tab:parchg}. In those plots it is possible to notice that both U and HSE were responsible for a better localization of these levels, which are clearly of Ti(d) character, presenting apparently the same hybridization with O(p) orbitals in both cases. This hybridization could be a problem, as pointed out in the literature, in the case of MnO polymorphs\cite{Peng2013}, where GW calculations where used as a benchmark for the correct ordering of Mn(d) levels. We could in principle follow the same path, but due to computational cost, we decided only to report those  defect levels, and point out that they could be responsible for n-doping in these structures, which could be a possible explanation for their enhanced conductivity on memristors. 

One possible interpretation for the differences between the PBESol+U and HSE results for the $\beta$-, $\gamma$-Ti${}_3$O${}_5$ and Ti${}_4$O${}_7$ is that both methods are known to present improvements to the electronic struture in this kind of system, but in different ways. The first uses the fact that the on-site Coulomb interaction (which in this case is inserted is the system through a Hubbard parameter) is responsible for the localization of Ti(d) electrons, which in turn leads to a better description of these orbitals. The hybrid approach relies on the fact that through the insertion of part of Hartree-Fock exchange in the exchange-correlation functional, there is a partial cancelation of the self interaction, considered as an intrinsic problem in GGA-based calculations\cite{Kim2009}. Those different approaches seem to play a decisive role in the final electronic structure of the systems being studied, as could be seen in the different total magnetizations for these two structures, but qualitatively their description of the position of the defect levels and their character is very similar. It is important to notice that the HSE calculations led to a metallic state for $\beta$- and $\gamma$-Ti${}_3$O${}_5$, while for all other structures, a semiconductor PDOS was obtained (it was possible to notice the separation between occupied and unoccupied levels). This same semiconducting behavior was obtained using the Hubbard U parameter. According to Rao \textit{et al}\cite{Rao197183} and Bartholomew and Frankl,\cite{Bartholomew1969} the systems studied in this work are all semiconductors for temperatures below 100 K. These results seem reasonable, given that DFT calculations always take place at T = 0 K.

In the case of Ti${}_2$O${}_3$, the magnetization is very weak and present only on PBESol calculations. When U was introduced, it becomes negligible, and the same physical interpretation can be given: better description of the localization of Ti(d) orbitals. The same behavior of U = 5 is seen when the HSE functional is used, with the exception of $\beta$-Ti${}_3$O${}_5$ and Ti${}_5$O${}_9$. This could be understood as observations of different local minima related to the different magnetic configurations. In fact, we performed other calculations for different magnetic orderings for all structures except Ti${}_5$O${}_9$ and a difference of few meV was found between the total energy per unit cell without any restriction on the magnetic moments (shown in Table \ref{tab:mags}) and restricting the initial configuration to antiferro - AF, $\mu = 0$ - configurations for Ti(d) electrons. This points out that those structures present a number of minima of the total energy with respect to magnetic ordering, but the difference is relatively small. In any case, it is clear that the localization is more pronounced in either situation.

% -----------------------------------------------------------------------------------------------------------

\section{Conclusion}

In this work we showed that DFT based calculations are able to point out defect levels inside the bandgap region for all Ti${}_n$O${}_{2n-1}$, $2 < n < 5$. Those defects, in the same way as defects on rutile TiO${}_2$, are mainly of Ti(d) character and exist because of the introduction of intrinsic V${}_{\mathrm{O}}$'s in the structure. Those levels are either close or attached to the CBM when GGA functionals are used for calculations, but a better description of Ti(d) orbitals - through the Hubbard U parameter, or a hybrid functional - can lead to the positioning of these levels away from the CBM, as shallow or even deep levels. Although we find an extended defect level - forming a narrow band - which has a local magnetic moment, the value of the exchange coupling J is very small (at least for the different configurations tried).

The enhancement of the electronic conductivity observed for these oxides with respect to rutile could be explained, in principle, by the presence of these defects close to the CBM, which play the role of intrinsic dopants in these systems. This fact should be important in order to better understand the memristive devices. Similarities between the electronic structures around the CBM and with respect to the deffect levels obtained with PBESol+U, U = 5 and HSE are remarkable for the Ti${}_2$O${}_3$, Ti${}_4$O${}_7$ and Ti${}_5$O${}_9$. The Ti${}_3$O${}_5$ on the other hand has a "metallic" character for HSE, while for PBESol+U, U = 5, a bandgap appeared.

 A comparison between the PBESol+U and HSE functionals was also presented, aiming to identify which of the two methodologies is best suited for the study of correlated oxide systems, as the Ti${}_n$O${}_{2n-1}$ Magn\'eli structures presented in this work. While the total magnetization results were not equal for all structures, the localization and character of the defect levels were similarly described with both methods. The fact that HSE calculations result in metallic behavior for two of the systems studied, which were all semiconductors according to PBESol+U calculations, while experimental evidence points out a semiconductor behavior, is an evidence that the latter methodology could be best suited for this kind of system.

% -----------------------------------------------------------------------------------------------------------

\begin{acknowledgments}
This work was supported by FAPESP, CNPq and CAPES (Brazil), and Vicerrectoría de Docencia-Universidad de Antioquia (Colombia). The support given by Cenapad-SP in the form of computational infrastructure is also acknowledged. The authors would like to thank Prof. Dr. H. Raebiger for the valuable advices.
\end{acknowledgments}

\end{document}